\documentclass[acmsmall,screen]{acmart}
\AtBeginDocument{%
  }

\acmISBN{978-1-4503-XXXX-X/2018/06}

\usepackage{adjustbox}
\usepackage{multirow}
\usepackage{graphicx}
\usepackage{stfloats}
\usepackage{caption}
\usepackage{subcaption}
\usepackage{enumitem}
\usepackage{makecell}

\makeatletter
\newcommand{\thickhline}{%
    \noalign {\ifnum 0=`}\fi \hrule height 1pt
    \futurelet \reserved@a \@xhline
}
\newcolumntype{"}{@{\hskip\tabcolsep\vrule width 1pt\hskip\tabcolsep}}
\makeatother
\definecolor{lightapricot}{rgb}{0.93, 0.92, 0.84}

\colorlet{lightgrey}{lightgray}
\usepackage{color,soul}
\usepackage[table]{xcolor}

\newcommand{\secref}[1]{Section~\ref{sec:#1}}
\newcommand{\tabref}[1]{Table~\ref{table:#1}}
\newcommand{\secsref}[2]{Sections~\ref{sec:#1}-\ref{sec:#2}}

\newcommand{\figref}[1]{Figure~\ref{fig:#1}}

\usepackage{tikz}

\newcolumntype{C}[1]{>{\centering\arraybackslash}p{#1}}

\begin{document}

\title{Hybrid LLM Routing for Efficient App Feedback Classification}

\author{Yasaman Abedini}
\orcid{0009-0005-8168-6116}
\affiliation{%
  \institution{Department
of Computer Engineering, Sharif University of Technology}
\country{Iran}
}
\email{y.abedini14@sharif.edu}

\author{Abbas Heydarnoori}
\orcid{0000-0001-9785-2880}
\affiliation{%
  \institution{Department
of Computer Engineering, Sharif University of Technology}
\country{Iran}}
\email{heydarnoori@sharif.edu}
\authornote{Abbas Heydarnoori is currently affiliated with the Department of Computer Science at Bowling Green State University, USA.}

\renewcommand{\shortauthors}{Abedini et al.}

\begin{abstract}
The emergence of large language models (LLMs), pre-trained on massive datasets, has demonstrated strong performance across a wide range of natural language processing (NLP) tasks, including text classification. While prior studies have examined the use of LLMs for predicting the intent of user feedback and reported encouraging results, these investigations remain limited in scope. Furthermore, the vast volume of feedback posted daily, particularly for popular applications, combined with the computational and financial overhead of commercial LLMs, renders large-scale deployment impractical. In contrast, smaller models provide greater efficiency and lower cost but generally at the expense of reduced accuracy. In this paper, we aim to balance accuracy and efficiency in feedback classification. We first present a comprehensive study of zero-shot classification using four widely adopted LLMs, GPT-3.5-Turbo, GPT-4o, Flan-T5, and Llama3-70B, on diverse feedback datasets collected from multiple platforms, including app stores, forums, and X, which are categorized under different schemes. This analysis reveals how classification scheme design and platform characteristics influence the predictive performance of LLMs. Building on these insights, we propose a two-tier routing strategy for scalable app store feedback classification. In this approach, low-complexity instances are processed by lightweight fine-tuned models, while ambiguous cases are routed to high-capacity LLMs for more reliable decisions. Experimental results show that this strategy retains 98.4\% to 100.4\% of zero-shot LLM accuracy while reducing request and token costs by 67.8\% and 66.3\%, respectively.

\end{abstract}



\received{11 September 2025}

\maketitle

\section{Introduction}
\label{sec:introduction}
Users can share and discuss their feedback with each other or with developers through various platforms, such as app stores, forums, issue tracking systems, and social media platforms~\cite{Devin, Nayebi, Tizard, Abedini}. These platforms serve as rich repositories of user requirements, and analyzing their content offers valuable insights that can aid developers in maintaining and improving their apps~\cite{Malgaonkar}.
In recent years, substantial research has conducted on identifying and classifying requirements extracted from these sources, yielding promising results~\cite{Iacob,Guzman,Gao,Villarroel,Stanik}.
Transformer-based models demonstrated state-of-the-art performance in feedback classification~\cite{Hadi}.
However, trained models fail to deliver the expected accuracy in prediction on unseen datasets~\cite{Devin}. This issue threatens the generalizability of these approaches and hinders their practical application~\cite{Dabrowski}.
Although efforts have been made to address this issue, it remains an open challenge that requires further research and investigation~\cite{Dabrowski,Devin,Abedini}.

In recent years, there has been a surge of interest in LLMs, especially variations of OpenAI's GPT~\cite{openai_models}, which has emerged as one of the most prominent and widely studied models in this domain~\cite{Liu}. These models have provided remarkable accuracy across various NLP tasks, including text classification, feature extraction, sentiment analysis, text generation, and data labeling~\cite{Liu,Hou,Ding}. Researchers leverage LLM-based techniques to develop agents that can address traditional NLP tasks with improved performance~\cite{Jin}.
Zhang et al.~\cite{Zhang} demonstrated that GPT-3.5 can effectively classify non-functional requirements and extract features in a zero-shot setting.
Gunathilaka et al.~\cite{Gunathilaka} conducted a limited study on 200 user feedback instances, highlighting the potential of LLMs in analyzing user feedback.
While these studies provide valuable preliminary insights, they are limited in scope and do not offer a comprehensive analysis of LLMs’ capabilities in predicting different feedback categories across diverse data sources.
Based on these findings, a key question arises: \emph{To what extent can LLMs provide accurate and efficient classification of user feedback?}

However, their practical use of LLMs faces notable challenges. LLMs are constrained by substantial computational requirements, response time latency, and significant operational costs. These limitations make large-scale adoption, such as processing the massive volumes of user feedback generated daily, particularly impractical.
Additionally, state-of-the-art fine-tuned transformer-based models are lighter and more efficient for processing large volumes of user feedback compared to LLMs, albeit at the lower accuracy.

Routing strategies have been proposed to improve the practicality of LLMs by balancing accuracy and efficiency. Building on this foundation, we propose a two-tier routing strategy specifically designed for user feedback classification. Our approach routes input samples to the most suitable models, leveraging the complementary strengths of state-of-the-art LLMs and fine-tuned models. This design directly addresses the challenges of accuracy, cost, and scalability in feedback classification, making it well-suited for real-world deployment. To the best of our knowledge, this is the first study to introduce a routing-based approach that explicitly targets efficient and scalable feedback classification.



In this paper, we first present a comprehensive empirical study evaluating LLMs in the zero-shot setting for user feedback classification. Specifically, we analyze feedback from eight datasets collected across three widely used platforms, employing various classification schemes with different levels of granularity. 
The datasets consist of four from Google Play~\cite{GPlay} and the App Store~\cite{AppStore}, two from the X platform~\cite{Twitter}, one from the VLC~\cite{VLC} and Firefox~\cite{Tizard} forums, and one from the Reddit~\cite{Reddit} forum. We leverage four diverse LLMs, GPT-3.5-turbo~\cite{GPT-3.5}\footnote{Hereafter referred to as GPT-3.5 throughout the paper.}, GPT-4o~\cite{GPT-4o}, Flan-T5~\cite{Flan-t5}, and Llama3-70b~\cite{Llama3}\footnote{Hereafter referred to as Llama3 throughout the paper.}.
Unlike prior studies~\cite{Zhang,Gunathilaka}, which have typically focused on limited datasets or specific model families, our work provides the first systematic and large-scale evaluation that spans diverse datasets and models in feedback calssification.
To evaluate performance, we conduct two experiments. In the first, we assess classification accuracy under the original schemes.
Our results show that LLMs perform better at classifying user feedback when the categories are clearly distinct. In contrast, applying LLMs to identify fine-grained categories often produces unsatisfactory results.
In the second experiment, we adapt the datasets to a coarse-grained classification scheme, following prior work~\cite{Devin,Abedini}. This scheme, which is particularly relevant to developers, defines three main categories: \emph{bug report}, \emph{feature request}, and \emph{other}, each with comprehensive definitions. Our results demonstrate that while LLMs face challenges in predicting fine-grained categories, they perform better when the categories are abstract, distinct, and well-defined.

Building on these insights, we propose a two-tier routing strategy designed to balance efficiency and accuracy. In this approach, feedback instances are first classified using fine-tuned models. Instances for which the fine-tuned models do not reach consensus are considered complex and are routed to high-capacity LLMs for final classification. To evaluate the effectiveness of our strategy, we compare its performance against a diverse set of baselines. These include fine-tuned models trained on existing manually labeled datasets as well as models augmented with general-purpose or app-specific datasets using the annotation capabilities of LLMs. 
Additionally, we assess both the \emph{request count} and \emph{token count} of our approach relative to these baselines. 
The results demonstrate that our routing strategy can retain 98.4\% to 100.4\% of this accuracy while reducing request and token costs by 67.8\% and 66.3\%, respectively. By effectively combining fine-tuned models and LLMs, the proposed approach provides a scalable and practical solution. This makes it well-suited for deployment in settings such as popular apps, where both accuracy and efficiency are critical. 




In general, this paper presents three main contributions:
\begin{enumerate}
    \item We conducted a comprehensive study to evaluate the capabilities of various LLMs in classifying user feedback from diverse sources following different classification schemes. Our findings demonstrate that selecting an appropriate classification scheme can enhance the accuracy of LLMs in identifying different types of user feedback. 
    
    \item We propose the first hybrid LLM routing strategy for user feedback classification that balances efficiency and accuracy by combining fine-tuned models with high-capacity LLMs. Results show that this approach significantly reduces costs while maintaining accuracy, providing a practical solution for large-scale applications.
    
    \item We have made our source code, collected data, and obtained results publicly available~\cite{replicationpackage} to support future research.

\end{enumerate}

\textbf{Paper Structure}: In~\secref{related_work}, we review relevant literature. In~\secref{problem_definition} we formulates our goal of this study.
In~\secref{rq1}, we introduce our experiments and analyze the obtained results for empirical analysis of zero-shot classification. In~\secref{proposed}, we describe our two-tier LLM routing strategy. In~\secref{rq2}, we present the experimental evaluation of our routing strategy. In~\secref{rq3}, we present the experimental evaluation of first-tier model configurations.
In~\secref{discussion}, we discuss the implications of our results.
In~\secref{threat} We examine potential threats to the validity of our experiments. Finally, we present our conclusions in~\secref{conclusion}.

\section{Related Work}
\label{sec:related_work}
In this section, we review prior research in two relevant areas: user feedback classification and LLM routing strategies.
\subsection{User Feedback Classification}
Extensive research has been conducted in user feedback classification, leveraging NLP techniques and machine learning algorithms with promising results. However, the lack of sufficient labeled dataset poses a significant challenge to the generalizability of these models~\cite{Devin,Dabrowski}. To address this issue, Devine et al.~\cite{Devin} demonstrated that integrating multiple labeled datasets from diverse sources can enhance model generalization. Additionally, Abedini et al.~\cite{Abedini} explored augmenting existing datasets with information extracted from labeled issues. Despite these advancements, current approaches still require further refinement to achieve optimal performance.

Recent studies have shown that using LLMs for information classification and data augmentation tasks has yielded promising results. He et al.~\cite{He} introduced an approach utilizing the GPT-3.5 model as an annotator, leveraging chain-of-thought prompts to enhance accuracy. Gilardi et al.~\cite{Gilardi} employed ChatGPT for five labeling tasks on a dataset from X, finding that the model outperformed crowd-workers. Moller et al.~\cite{Moller_1} generated new data using GPT-series models to augment datasets with limited samples; however, in two out of three tasks, the generated data did not yield satisfactory results. Ding et al.~\cite{Ding} used the GPT-3 model for both data generation and labeling, conducting experiments on text classification and named entity recognition tasks. Colavito et al.~\cite{Colavito} focused on issue labeling, demonstrating that GPT-series models could achieve satisfactory results in a zero-shot setting. Given these studies’ demonstration of LLMs achieving competitive results with state-of-the-art approaches, further research is needed to investigate how these models can effectively address the challenges of user feedback classification.

Zhang et al.~\cite{Zhang} introduced a framework for evaluating ChatGPT's capabilities across four tasks related to information classification and extraction. In one task, they assessed the model's performance in classifying a dataset of non-functional app feedback in a zero-shot setting. Nevertheless, the study offers limited insights into feedback classification and does not provide a comprehensive examination of how different LLMs tackle its associated challenges. To the best of our knowledge, no prior research has addressed this topic.
In this research, we analyze the capabilities of four LLMs in classifying feedback across eight datasets from three platforms, each following diverse classification schemes at varying levels of abstraction. By evaluating these models’ labeling capabilities against state-of-the-art approaches, we aim to conduct a comprehensive study that bridges the existing research gap in this field.

\subsection{LLM Routing Strategies}
Routing strategies in LLM systems aim to balance accuracy and efficiency by selecting between models of varying capacities and costs. In this approach train router models that estimate query difficulty and dynamically assign tasks. For example, RouteLLM demonstrates more than a twofold reduction in cost without sacrificing response quality~\cite{Ong}. Building on this idea, recent work has introduced more advanced architectures, such as PolyRouter, which balances performance and cost, and workload-aware schedulers that maximize throughput under high computational demand~\cite{Zhao_1}.
More sophisticated routing methods go beyond binary decision-making and incorporate explicit optimization of cost–accuracy trade-offs. Cost-aware frameworks like CARROT use principled statistical approaches to jointly model performance and cost, with minimax analysis showing that routers considering both dimensions can achieve optimal results~\cite{Somerstep}. Similarly, OmniRouter formulates routing as a constrained optimization problem, minimizing cost while maintaining required performance levels~\cite{Mei_2}.

In contrast to existing general-purpose routing systems, which are primarily grounded in standard benchmarks, our study presents a routing strategy specifically designed for app feedback classification. By accounting for the characteristics of available datasets and the intrinsic properties of user feedback, the proposed approach provides a practical solution that effectively balances efficiency and accuracy in real-world applications.


\section{Problem Definition}
\label{sec:problem_definition}
With thousands of similar apps vying for user attention, developers face intense competition to attract and retain users~\cite{noei}. Effective identifying and understanding user needs and preferences and addressing existing app issues are essential for developers to succeed in this competition~\cite{Gu}. Due to the large volume of submitted feedback, numerous approaches have been proposed for automatic classification of them, which in turn facilitates systematic analysis~\cite{Dabrowski,Hadi,Abedini}. While prior research indicates that LLMs can enhance analytical accuracy~\cite{Colavito,Gunathilaka}, their application is associated with three key challenges: 

\emph{Dataset Heterogeneity}: Different platforms (e.g. app stores, user forums) attract distinct groups of users depending on their goals and functional nature. This causes the feedback posted on these platforms to follow various writing styles, including variations in vocabulary, structure, and tone. Consequently, applicability of generalizable approach using LLMs faces certain limitations. Understanding the effectiveness of existing approaches in analyzing feedback across different platforms is an important aspect of problem definition.

\emph{Diversity of Classification Schemes}: Researchers analyze user feedback from different perspectives and at varying levels of granularity. The complexity and overlap among defined categories can significantly influence the predictability of LLMs. Therefore, adopting a classification scheme that aligns with the capabilities of LLMs is crucial for efficient feedback analysis.

\emph{Efficiency–Accuracy Trade-off}: While high-capacity LLMs often achieve superior zero-shot accuracy, their computational cost and latency limit their practical deployment. Conversely, smaller models or simpler heuristics are more efficient but often suffer from reduced accuracy. An effective solution must therefore balance these competing objectives.

Building on these challenges, we first design a comprehensive empirical study to examine the zero-shot classification of user feedback across multiple platforms 
and using various classification schemes with different levels of granularity. 
This study allows us to understand the conditions under which LLMs perform 
effectively and to identify the limitations that motivate a more efficient routing strategy.

Considering these limitations, we formalize the problem as follows: given a user feedback instance $f$, a classification scheme $S$, and an available set of models  $\{M_1, M_2, \dots, M_k\}$ with different capacities and costs, the goal is to assign a label $y \in S$ such that: 

\[
\text{maximize} \quad
\mathcal{P}(M, f, S) \quad
\text{while minimizing} \quad \mathcal{C}(M, f).
\]

In this paper, the predictive performance of a classification task is denoted by $\mathcal{P}$. Depending on the experimental setting, $\mathcal{P}$ is evaluated using standard classification metrics, including precision, recall, and F1-score.
The concept of $\mathcal{C}$ is examined along two dimensions. 
The first, \emph{request count}, refers to the number of model 
invocations; a higher number of requests increases the overall processing time. 
We avoid using raw latency as a measure, since it depends on external factors 
such as network conditions and server load. 
The second, \emph{token count}, refers to the cumulative number of tokens transmitted across all requests, which directly affects the monetary expense for commercial LLMs such as the GPT series. 

To address this problem, we propose a two-tier LLM routing strategy that balances efficiency and accuracy by leveraging smaller models for simple cases and high-capacity LLMs for ambiguous ones. Accordingly, we investigate the following research questions: 

\begin{itemize}
    \item \textbf{RQ1:} To what extent does the accuracy of zero-shot classification differ across various classification schemes and diverse feedback platforms?
    \item \textbf{RQ2:} To what extent does the proposed two-tier LLM routing strategy enhance efficiency while preserving accuracy, compared to state-of-the-art feedback classification approaches?
    \item \textbf{RQ3: } How do different first-tier configurations influence the efficiency and accuracy of the proposed two-tier LLM routing strategy?
\end{itemize}

\section{Empirical Analysis of Zero-Shot Classification (RQ1)}
\label{sec:rq1}
To address our first research question, we designed two experiments to evaluate the ability of LLMs to classify different types of user feedback under classification schemes with varying levels of abstraction. For this purpose, we utilized eight manually labeled datasets from prior studies, drawn from multiple publicly available sources: four from app stores, two from the X platform, and two from online forums. For comparability and generalizable results, we conducted experiments with four widely adopted LLMs, including both commercial and open-source models. We also designed carefully appropriate prompts that explicitly included category definitions to guide the models. The remainder of this section is organized as follows. We first describe the experimental setup (\secref{experimental_setup}) and then present the two experiments in detail (\secsref{rq1_experiment1}{rq1_experiment2}).


\subsection{Experimental Setup}
\label{sec:experimental_setup}
\subsubsection{Dataset Selection}
\label{sec:select-dataset}
In recent years, researchers have analyzed user feedback from multiple perspectives and proposed various classification schemes tailored to different objectives~\cite{Hadi}. To ensure a comprehensive evaluation, we selected manually labeled datasets according to two criteria: (i) the use of rigorous labeling processes, and (ii) coverage of diverse sources employing different classification schemes. Accordingly, we chose datasets extracted from three common user feedback sources.~\tabref{Adopt_Labels} provides detailed information about the datasets used in this study.

\textbf{App Store Datasets:}
App stores are among the most widely used platforms for sharing user feedback, providing valuable insights across a broad range of applications. As a result, a substantial portion of past research has primarily focused on analyzing data extracted from app stores~\cite{Dabrowski}. In this study, we adopt four manually labeled datasets that have also been used in earlier empirical studies~\cite{Devin,Hadi}. These datasets were collected from Google Play and the App Store platforms, encompassing user feedback from a variety of applications. Since a subset of DS3 overlapped with DS2, we removed the duplicate instances to avoid redundancy and potential bias in the results~\cite{Abedini}.

\textbf{Forum Datasets:}
Online forums dedicated to applications represent another valuable source of user feedback, often helping developers identify requirements and issues~\cite{Tizard, Iqbal}. These forums typically host active communities of users and developers who engage in diverse, interactive discussions to exchange support and guidance. In this study, we use two datasets, DS5 and DS6, collected from the Reddit, VLC, and Firefox forums.

\textbf{X Datasets:}
Platform X (formerly Twitter) is another important source of user feedback on applications~\cite{Guzman_2}. Prior research has shown that analyzing tweets can complement insights gained from app store feedback~\cite{Nayebi}. Although tweets are typically shorter than app store feedback, they still convey valuable information for developers~\cite{Guzman_2}. Ignoring them may lead to missing critical insights. In this study, we employ two manually labeled tweet datasets, DS7 and DS8.

To ensure more accurate analyses, we filtered out samples that lacked sufficient information. Specifically, only instances containing at least three words after standard preprocessing, such as lowercasing, and the removal of numbers, punctuation, stopwords, and non-informative characters (e.g., \$, \#), were retained~\cite{Assi}.

\begin{table*}
    \caption{Details of the target manually labeled datasets in our study}
    \centering
    \begin{tabular}{|l|l|l|p{.6\linewidth}|}
        \hline
        \textbf{Platform} & 
        \textbf{Dataset} & 
        \textbf{\#Samples} & 
        \textbf{Category Names}
        \\
        \hline
        \multirow{4}{*}{App Stores}
        &
        DS1~\cite{Guzman} & 3,887 & bug report, user request, praise, usage scenario\\
        \cline{2-4}
        &
        DS2~\cite{Maalej_Journal} & 2,725 & bug report, feature request, user experience, rating\\
        \cline{2-4}
        &
        DS3~\cite{Jha} & 3,214 & bug report, feature request, other\\
        \cline{2-4}
        &
        DS4~\cite{Scalabrino} & 2,826 & functional bug report, suggestion for new feature, other\\
        \hline
        \multirow{4}{*}{Forums} &
        DS5~\cite{Tizard} & 2,997 & apparent bug, feature request, usage, none-informative, application guidance, question on application, help seeking, user setup\\
        \cline{2-4}
        &
        DS6~\cite{Iqbal} & 1,731 & bug report, user requirement, miscellaneous and spam\\
        \hline
        \multirow{2}{*}{X} &
        DS7~\cite{Williams} & 3,839 & bug, feature, other\\
        \cline{2-4}
        &
        DS8~\cite{Stanik} & 9,276 & problem report, inquiry, irrelevant\\
        \hline
    \end{tabular}
\label{table:Adopt_Labels}
\end{table*}

\subsubsection{LLM Selection}
To evaluate the capabilities of LLMs, we selected two closed-source models, GPT-3.5-turbo\cite{GPT-3.5} and GPT-4o\cite{GPT-4o}, and two open-source models, Llama3\cite{Llama3} and Flan-T5\cite{Flan-t5}. These models were chosen based on their applicability during our study and their established performance across various tasks in previous studies~\cite{Dvivedi,Ye}. 
The first two models are recent releases from OpenAI's GPT-3.5 and GPT-4 series, known for their high accuracy and fast response times~\cite{GPT-3.5-Turbo-Updates,hello-gpt-4o}. GPT-3.5-turbo is optimized for chat and demonstrates better instruction understanding than earlier GPT-3 models while offering lower costs~\cite{Ye}. GPT-4o enhances the capabilities of its predecessors and provides higher speed~\cite{Introducing-GPT-4o}. It achieves performance comparable to GPT-4 Turbo in various text benchmarks~\cite{hello-gpt-4o}. As a result, they can be suitable candidates for conducting our designed experiments.

Llama3, introduced by Meta~\cite{meta-llama-3}, has demonstrated competitive results with GPT-4o across various tasks~\cite{Dubey}. Comparing its capabilities with GPT-series models in classifying user feedback can help developers select the most suitable model for their specific objectives. In this study, we utilize a pre-trained Llama3 model with 70 billion parameters~\cite{Llama3}. Moreover, the three previously considered models in this study follow a decoder-only architecture. To ensure a more comprehensive evaluation, we also include Flan-T5~\cite{Flan-t5}, an encoder-decoder model, as one of our target models. Flan-T5 enhances the performance of the T5 model through instruction fine-tuning~\cite{Chung}, making it a valuable addition to our analysis.

To interact with the three LLMs, GPT-3.5, GPT-4o, and Llama3, we submitted individual requests for each sample through the APIs of their respective providers. The Flan-T5 model was executed locally on a free GPU available via the Google Colab platform. While some model outputs explicitly returned the predicted category, in many cases additional post-processing was required to extract the relevant labels. To address this, we implemented an automated function to parse and extract the predicted class from the raw outputs. In cases where multiple categories were produced, manual verification was performed to ensure the correctness of the extracted classes. Further details regarding hyperparameter values and post-processing steps are provided in our shared artifacts.

\subsubsection{Prompt Preparation}
In zero-shot learning, the quality of the prompt is critical for guiding the model to generate accurate and relevant responses. A well-structured prompt clearly defines the task, provides essential context, and specifies the desired output format~\cite{Colavito}. Since prompt design directly affects model performance, we carefully constructed task-specific prompts following insights from prior research~\cite{Colavito}. 

\begin{figure}
\centering
  \includegraphics[width=4in,height=2.5in]{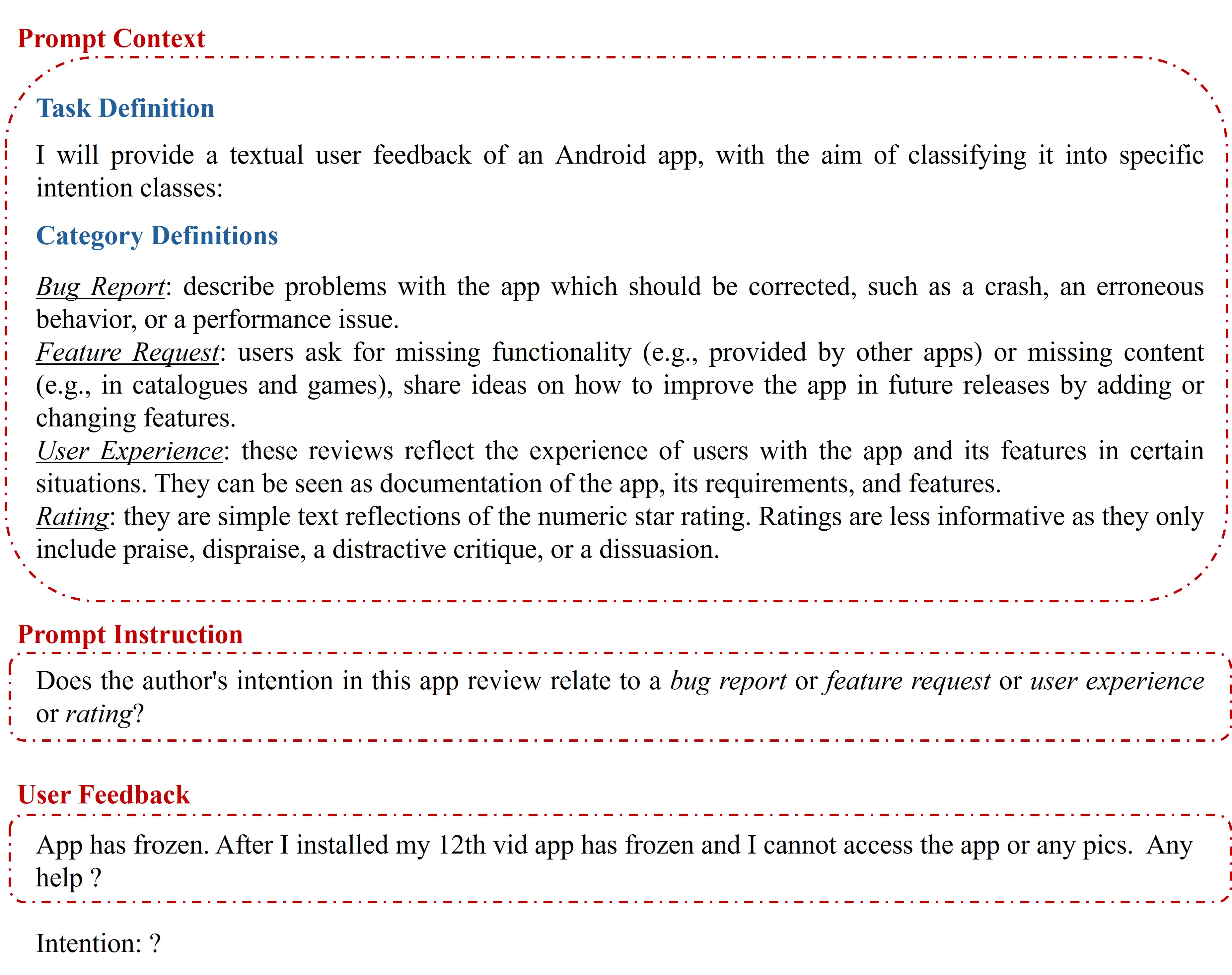}
\caption{Sample prompt for zero-shot classification of user feedback in dataset DS2}

\label{fig:sample_prompt}  
\end{figure}

Each prompt used in our experiments consists of two main components: \emph{prompt context} and \emph{prompt instruction}. The \emph{context section} provides background information, including a description of the classification task and an introduction to the target categories. The \emph{instruction section} explicitly asks the model to classify the given user feedback into the most appropriate category based on the provided context. To align with the objectives of each experiment, we designed tailored prompts that incorporate relevant task details and category definitions. For instance,~\figref{sample_prompt} illustrates the prompt used for classifying DS2 under its original scheme. The exact prompts used in each experiment are described in the corresponding sections.

\subsection{Experiment 1: Classification with Original Schemes}
\label{sec:rq1_experiment1}
This experiment evaluates LLM performance in classifying user feedback using the original schemes of eight manually labeled datasets (DS1–DS8) with varying levels of granularity, under a zero-shot setting. For each dataset, we constructed prompts that included the descriptions of the predefined categories as reported in the corresponding original studies. As a result, conceptually similar categories may appear under different names and with varying definitions across datasets. For instance, the category related to bug reports is present in all datasets but labeled differently: \emph{bug report} in DS1, DS3, and DS6; \emph{bug} in DS2 and DS7; \emph{functional bug report} in DS4; \emph{apparent bug} in DS5; and \emph{problem report} in DS8. The complete set of prompts for all datasets is available in the shared artifacts for reproducibility. 


For comparison, we employ Transformer-based models, which have demonstrated state-of-the-art performance in feedback classification~\cite{Hadi}, as baselines. We fine-tune binary classification models for the defined categories using the pre-trained DistilBERT model~\cite{Distilbert_model}, implemented via the \emph{AutoModel} function~\cite{TFAutoModel} from the Hugging Face library. To ensure robust evaluation for categories with fewer samples, we adopt 5-fold cross-validation. Class imbalance is addressed by incorporating category-specific weights during training through the \emph{class\_weight} parameter.

\subsubsection{Results}
~\tabref{RQ1_results} presents the evaluation results of LLM predictions. For clarity, the maximum and minimum values for each metric in each row are highlighted in bold and underlined, respectively. As shown, applying various LLMs in a zero-shot setting yields overall average improvements in precision ranging from 7.7 to 11.7, recall from 0.8 to 4.9, and F1 from 0.5 to 12.8, compared to the fine-tuned baseline model. The notable improvement in F1 is primarily attributable to a substantial increase in precision, while recall shows only modest improvement, with a decline observed for the Flan-T5 model. This indicates that feedback classification by the target LLMs does not necessarily ensure a higher likelihood of identifying all instances within a category. A potential explanation lies in the variation of feedback labeling policies among the annotators of different datasets. Such variations may enable models fine-tuned on the same dataset to identify a broader range of instances with probabilities comparable to those of stronger LLMs. Since LLMs are general-purpose models and not specifically fine-tuned for user feedback classification, significant improvements in identifying all types of feedback instances cannot be guaranteed.

\begin{table*}
    \caption{Results of feedback classification using target LLMs based on the primary scheme in zero-shot setting for RQ1}
    \centering
    \setlength{\tabcolsep}{3pt}
    \makebox[\textwidth][c]{
    \begin{tabular}{|p{.02\linewidth}|l|ccc|ccc|ccc|ccc|ccc|}
        \hline       
            \multirow{2}{*}{\textbf{}}
            & 
           \multirow{2}{*}{\textbf{Dataset}}
            &  \multicolumn{3}{|c|}{\textbf{GPT-3.5}}
            &  \multicolumn{3}{|c|}{\textbf{GPT-4o}} 
            &  \multicolumn{3}{|c|}{\textbf{Flan-T5}} 
            & \multicolumn{3}{|c|}{\textbf{Llama3}} 
            & \multicolumn{3}{|c|}{\textbf{Fine-Tuned}} \\
            \cline{3-17}
            & &
            P & R & F1 &
            P & R & F1 &
            P & R & F1 & 
            P & R & F1 &
            P & R & F1 \\
        \hline
        \parbox[t]{1mm}{\multirow{5}{*}{\rotatebox[origin=c]{90}{App Stores}}}
        &
        DS1 &
        \textbf{67.6} & 46.5 & \textbf{47.2} &
        59.7 & 50.2 & \textbf{47.2} &
        55.4 & \underline{39.7} & 43.2 & 
        63.4 & 52.3 & 47.1 &
        \underline{44.8} & \textbf{67.8} & \underline{40.3}
        \\
        \cline{2-17}
        &
        DS2 &
    	\textbf{59.9} & 57.2 & 57.4 &
    	58.7 & 61.5 & \textbf{58.3} &
    	51.0 & \underline{43.2} & \underline{28.4} &
    	52.0 & 67.1 & 57.6 &
    	\underline{46.8} & \textbf{71.7} & 51.8
    	\\
    	\cline{2-17}
        &
        DS3 &
	    80.6 & 77.2 & 77.5 & 
	    84.6 & 80.7 & \textbf{82.0} &
	    \textbf{85.2} & \underline{69.9} & 73.8 &
	    82.7 & 79.0 & 79.6 & 
	    \underline{60.7} & \textbf{81.1} & \underline{63.8}
        \\
    	\cline{2-17}
         &
        DS4 &
	    \textbf{73.2} & 73.5 & 69.9 & 
	    72.1 & 81.0 & \textbf{75.8} &
	    70.8 & \underline{53.9} & \underline{54.7} & 
	    59.0 & 74.8 & 56.0 & 
	    \underline{55.5} & \textbf{85.1} & 59.2
	    \\
	    \cline{2-17}
	    &
	    Average &
	    \textbf{70.3} & 63.6 & 63.0 & 
	    68.8 & 68.4 & \textbf{65.8} &
	    
	    65.6 & \underline{51.7} & \underline{50.0} & 
	    
	    64.3 & 68.3 & 60.1 &
	    
	    \underline{52.0} & \textbf{76.4} & 53.8
        \\
        \hline
        \parbox[t]{1mm}{\multirow{3}{*}{\rotatebox[origin=c]{90}{Forums}}}  
        &
        DS5 &
        \textbf{39.9} & 67.9 & 41.6 & 
    	37.7 & 69.5 & \textbf{43.8} & 
    	39.0 & 57.2 & 34.9 & 
    	41.3 & \textbf{71.4} & 47.9 & 
    	\underline{27.1} & \underline{30.4} & \underline{20.6}
    	\\
        \cline{2-17}
        &
        DS6 &
    	70.7 & 70.3 & 69.3 &
    	\textbf{71.2} & 70.3 & 70.7 & 
    	\underline{65.8} & \underline{61.7} & \underline{57.9} & 
    	\textbf{71.2} & \textbf{74.2} & \textbf{72.2} &  
    	67.6 & 55.0 & 53.0
    	\\
    	\cline{2-17}
    	& 
    	Average & 
	    55.3 & 69.1 & 55.5 & 
	    54.5 & 69.9 & 57.3 & 
	    52.4 & 59.5 & 46.4 &
	    \textbf{56.3} & \textbf{72.8} & \textbf{60.1} & 
	    \underline{47.4} & \underline{42.7} & \underline{36.8}
	    \\
	    \hline
         \parbox[t]{1mm}{\multirow{3}{*}{\rotatebox[origin=c]{90}{X}}}   
        &
        DS7 &
        76.0 & 77.5 & 76.6 & 
    	\textbf{80.2} & \textbf{80.4} & \textbf{80.3} & 
    	72.3 & 72.9 & 70.4 & 
    	77.9 & 80.0 & 78.6 & 
    	\underline{70.6} & \underline{69.0} & \underline{66.5}
        \\
        \cline{2-17}
        &
        DS8 &
	    55.6 & 54.6 & 50.1 & 
	    58.5 & \textbf{60.5} & \textbf{54.0} & 
	    \underline{51.8} & 53.3 & 50.4 & 
	    \textbf{58.6} & 58.9 & 50.5 & 
	    53.2 & \underline{43.9} & \underline{40.5}
        \\
        \cline{2-17}
        & 
        Average & 
    	65.8 & 66.1 & 63.4 & 
    	\textbf{69.4} & \textbf{70.5} & \textbf{67.2} & 
    	62.1 & 63.1 & 60.4 & 
    	68.3 & 69.5 & 64.6 & 
    	\underline{61.9} & \underline{56.5} & \underline{53.5}
        \\
        \hline
        \cline{1-17}
        -
        & 
        Overall Average & 
        \textbf{65.4} & 65.6 & 61.2 & 
        65.3 & 69.3 & \textbf{64.0} & 
        61.4 & \underline{56.5} & 51.7 & 
        63.3 & \textbf{69.7} & 61.2 & 
        \underline{53.7} & 64.8 & \underline{51.2}
        \\
        \hline
    \end{tabular}
    }
\label{table:RQ1_results}
\end{table*}

The comparative analysis of the four LLMs further reveals that GPT-4o consistently outperformed the others, achieving an average F1 of 64.0 and ranking highest across all datasets except DS6. GPT-3.5 and Llama3 achieved competitive results, both with an average F1 of 61.2. In contrast, Flan-T5 lagged behind, with an average F1 of 51.7, ranking lowest among the LLMs except in DS8. The superior performance of GPT-4o, GPT-3.5, and Llama3 may be attributed to their richer and more diverse training data, which likely enhances their ability to generalize across user feedback classification task.

Across all datasets, the use of GPT-3.5, GPT-4o, and Llama3 for predicting user feedback categories led to higher precision and F1 compared to the baseline. While Flan-T5’s predictions generally followed this trend for these metrics, decreases were observed in two datasets for precision and three datasets for F1. Conversely, the recall results for the LLMs exhibited variation, particularly in app store user review datasets, where the average recall of LLM predictions decreased by 8.0 to 24.7 compared to the baseline model. Therefore, while LLMs achieved higher precision for app store feedback, they did not consistently identify a greater proportion of instances within each target category. 
The results for the forum and X datasets indicate that, with only a few exceptions, the use of LLMs improved performance across all three evaluation metrics. 

The maximum improvements in average F1 achieved by LLMs were 23.3 for forums, 13.5 for X, and 12.0 for app stores. A cross-platform comparison of the average metrics shows that the relative improvements over the fine-tuned model were most pronounced for the forum datasets. This outcome is expected, as the forum datasets contained fewer samples per class (average 298.8) compared to the App Store (average 670.0) and X (average 2205.5). The smaller sample size in forums limited the effectiveness of the fine-tuned model, thereby amplifying the relative gains obtained by LLMs.

Moreover, the results show that the highest F1 scores in the zero-shot setting are frequently observed in DS7 and DS3. This stronger performance can be attributed to their coarse-grained classification schemes and the availability of relatively comprehensive category definitions, which help reduce ambiguity. 
The~\tabref{rq1_category_definition} presents these definitions, and the complete set of category definitions for all datasets is provided in the shared artifacts. By contrast, some datasets provide less detailed definitions. For instance, in DS1 the categories \emph{bug report} and \emph{feature shortcoming} are conceptually close, and their definitions are as follows:

\begin{itemize}
    \item \emph{Bug report: reviews that report a problem, such as faulty behavior of the application or of a specific feature.}
    \item \emph{Feature shortcoming: reviews that identify an aspect of an existing feature that users are unsatisfied with.}
\end{itemize}

These definitions provide limited distinction and require additional clarifications and examples to better disambiguate borderline cases. Such clarity is particularly important in zero-shot settings, where well-defined categories enable models to better capture conceptual boundaries and thus improve classification accuracy.

\begin{table*}
    \caption{Category definitions provided to LLMs for DS3 and DS7 datasets}
    \centering
    \makebox[\textwidth][c]{
    \begin{tabular}{|l|l|p{.8\linewidth}|}
        \hline       
        \textbf{DS\#} & 
        \textbf{Category} &  \textbf{Definition}
        \\
        \hline
        \multirow{3}{*}{DS3} &
        Bug Report & 
        Describe problems with the app which should be corrected, such as a crash, an erroneous behavior, or a performance issue.
        \\
        \cline{2-3}
        &
        Feature Request	& 
        Users ask for missing functionality (e.g., provided by other apps) or missing content (e.g., in catalogues and games), share ideas on how to improve the app in future releases by adding or changing features.
        \\
        \cline{2-3}
        &
        Other &
        Reviews that do not belong to any other categories.
        \\
        \hline
        \multirow{3}{*}{DS7} &
        Bug &
        All posts that report on bugs and errors of a software application.
        \\
        \cline{2-3}
        &
        Feature &
        All posts that contain information about the feedback on a feature (e.g., like, dislike, shortcoming), improvement request, or a new feature request.
        \\
        \cline{2-3}
        &
        Other &
        All posts that contain non-technical information related to software applications.
	    \\
        \hline
        
    \end{tabular}
    }
\label{table:rq1_category_definition}
\end{table*}

\begin{figure*}[!t]
    \centering
    \captionsetup[subfloat]{labelfont=footnotesize ,textfont=footnotesize}
    \subfloat[Misclassification heatmap for DS1]{\includegraphics[width=2.5in]{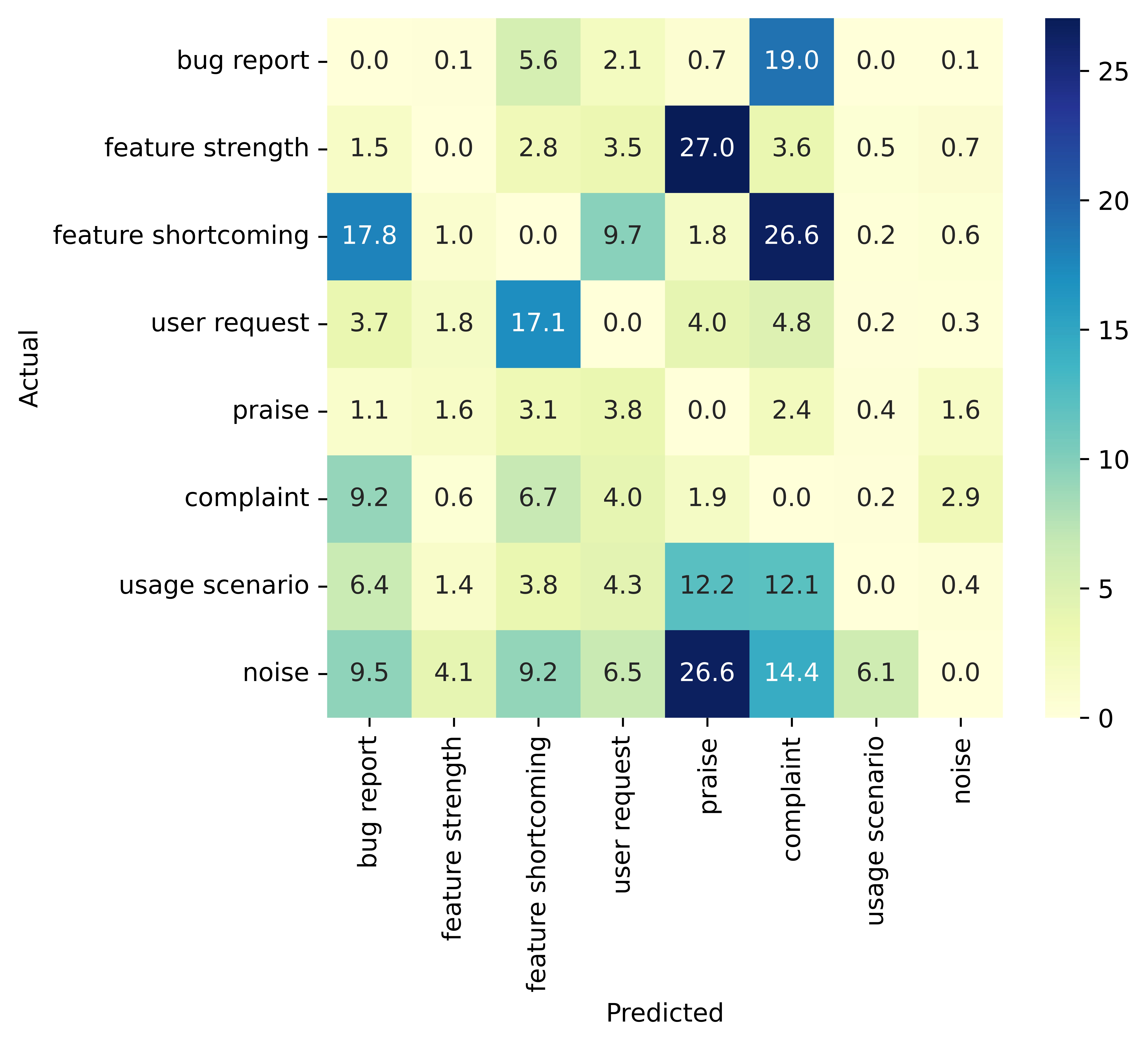}
    \label{fig:rq1_heatmap_ds1}}
    \hfil
    \subfloat[Misclassification heatmap for DS5]{\includegraphics[width=2.5in]{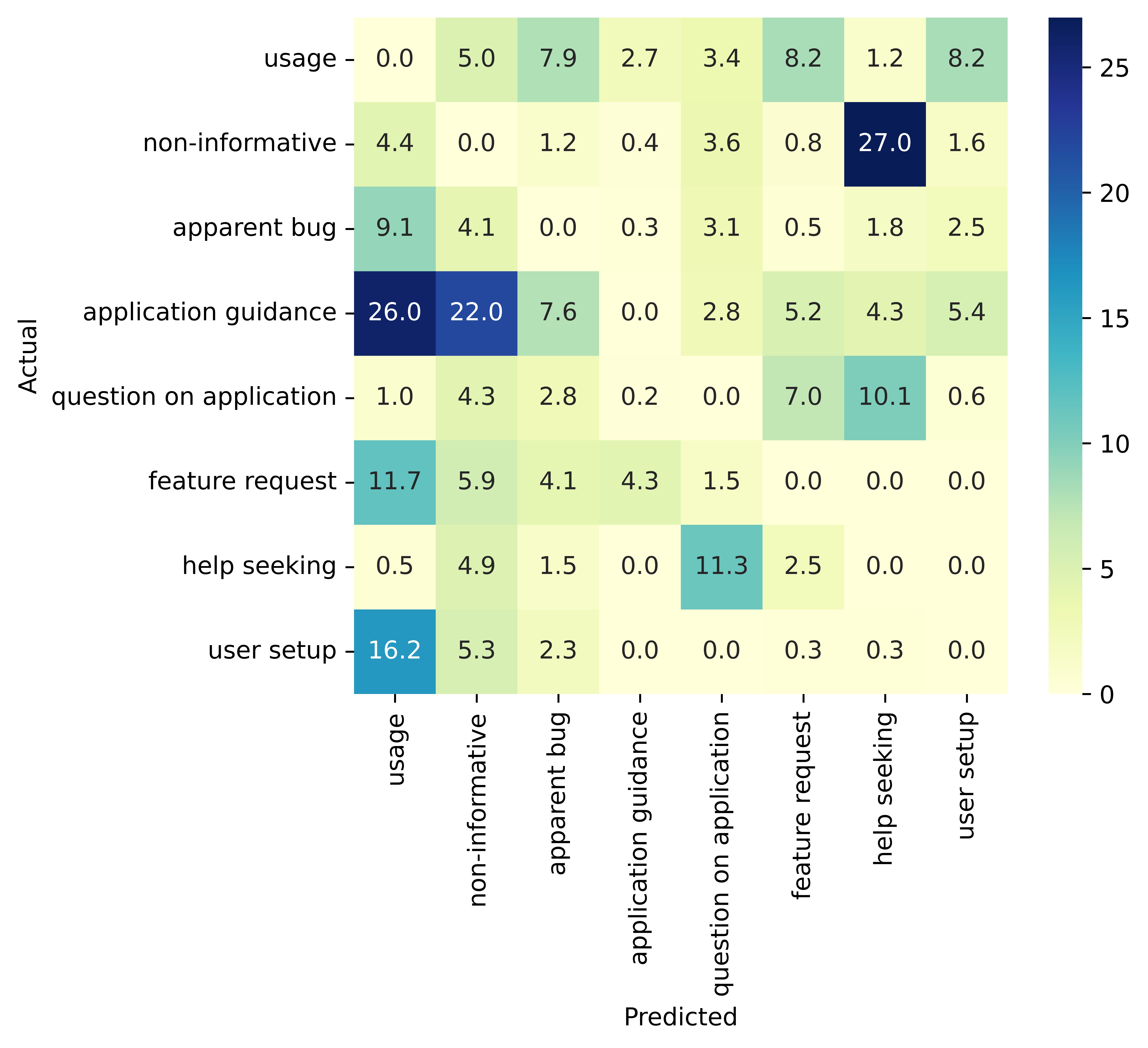}
    \label{fig:rq1_heatmap_ds5}}
    \caption{Cross-category misclassification percentages in DS1 and DS5}
\label{fig:rq1_heatmaps}
\end{figure*}

The lowest F1 scores in the zero-shot setting are observed in DS5 and DS1, which employ finer-grained schemes that contain some closely related categories. This increases the difficulty of distinguishing between samples and leads to more classification errors.~\figref{rq1_heatmaps} presents heatmaps for DS1 and DS5 illustrating misclassifications. Misclassification is defined as the proportion of instances assigned to an actual category $\mathcal{X}$ that were incorrectly predicted as belonging to category $\mathcal{Y}$, where $\mathcal{Y}$ is not included among their true labels. The reported values represent averages across the four evaluated LLMs.

As shown in~\figref{rq1_heatmap_ds1}, in DS1, 17.8\% and 26.6\% of samples labeled under the \emph{feature shortcoming} category were misclassified as \emph{bug report} and \emph{complaint}, respectively, both conceptually close categories. Likewise, 27\% of samples under\emph{feature strength} were misclassified as \emph{praise}, reflecting the inherent overlap between these categories, since user satisfaction often involves both highlighting strengths and expressing appreciation. Similar challenges appear in DS5 (~\figref{rq1_heatmap_ds5}), where several categories are conceptually close. For instance, 26.0\% of samples in \emph{application guidance}, 11.7\% in \emph{feature request}, and 16.2\% in \emph{user setup} were misclassified as \emph{usage}. Such confusion underscores how ambiguous or overlapping category definitions can lead to misclassification, particularly when models have not been explicitly trained on datasets structured around such nuanced distinctions. This highlights the importance of designing clearly defined, distinct categories to improve classification accuracy, especially in zero-shot settings.

\subsection{Experiment 2: Classification with Coarse-Grained Scheme}
\label{sec:rq1_experiment2}

In our study, various datasets have proposed diverse schemes, ranging from fine-grained taxonomies with eight categories~\cite{Guzman} to higher-level schemes with three classes~\cite{Jha}. Although such taxonomies are often tailored to specific purposes, our analysis indicates that fine-grained schemes with closely related categories tend to reduce the prediction accuracy of LLMs in zero-shot settings, thereby limiting their practical applicability. We designed the second experiment to assess whether employing a well-defined, coarse-grained classification scheme with distinct categories enables LLMs to better leverage their capabilities by reducing ambiguity.

To address this issue, we adopted a coarse-grained scheme consisting of three categories commonly of interest to developers: \emph{bug report}, \emph{feature request}, and \emph{other}. These categories are central to requirements management and have been widely examined in prior work~\cite{Devin,Hadi,Abedini}. Following earlier studies~\cite{Devin,Abedini}, we mapped the categories of all eight datasets (DS1–DS8) to this unified three-class scheme. The detailed mapping is provided in~\tabref{rq1_new_scheme_details}. Additionally, we constructed comprehensive definitions for each category by consolidating the descriptions used in Experiment 1. These unified definitions articulate the purpose of each category and emphasize distinctions based on common insights across datasets, thereby helping LLMs more accurately infer user intent. The complete definitions are presented as follows:

\begin{itemize}
    \item \emph{Bug Report}: Feedback that report a problem, such as faulty behavior of the application. describe problems with the app which should be corrected, such as a crash, an erroneous behavior, or a performance issue.
    \item \emph{Feature Request}: Feedback that ask for a missing feature, functionality, or content, as well as feedback that ask for the improvement of an existing feature
    \item \emph{Other}: Feedback where users express general appreciation with the application. It focuses on general judgment. They are simple text reflections of the numeric star rating. Ratings are less informative as they only include praise, dispraise, a distractive critique, or a dissuasion. Feedback where users describe workarounds, use cases and scenarios involving the app.
\end{itemize}

Results from the corresponding categories in the original classification schemes, mapped to the coarse-grained scheme, are used as baselines to facilitate a thorough comparison of LLM performance.

\begin{table*}
    \caption{Mapping of categories from DS1–DS8 to the coarse-grained classification scheme}
    \centering
    \begin{tabular}{|l|p{.22\linewidth}|p{.27\linewidth}|p{.4\linewidth}|}
        \hline
        \textbf{DS\#} & 
        \textbf{Bug Report} & 
        \textbf{Feature Request} & 
        \textbf{Other}
        \\
        \hline
        DS1 &
        bug report & user request & praise, usage scenario\\
        \hline
        DS2 &
        bug report & feature request & user experience, rating\\
        \hline
        DS3 &
        bug report & feature request & other\\
        \hline
        DS4 &
        functional bug report & suggestion for new feature & other\\
        \hline
        DS5 &
        apparent bug & feature request & usage, none-informative, application guidance, question on application, help seeking, user setup\\
        \hline
        DS6 &
        bug report & user requirement & miscellaneous and spam\\
        \hline
        DS7 &
        bug & feature & other\\
        \hline
        DS8 &
        problem report & inquiry & irrelevant\\
        \hline
    \end{tabular}
\label{table:rq1_new_scheme_details}
\end{table*}

\subsubsection{Results}
\figref{rq1_shaded_figures} presents a comparative analysis of LLM results under the original and coarse-grained classification schemes, highlighting the ranges of precision, recall, and F1-scores across datasets for the four LLMs. As observed, the use of a coarse-grained classification scheme improved the accuracy of feedback category prediction in the zero-shot setting in most cases. Specifically, predictions under the coarse-grained scheme showed the highest gains for the App Store dataset, with average increases of 10.1 in F1 and 9.8 in recall. The forum datasets also improved, with average increases of 4.6 in F1 and 6.2 in recall. Overall, F1 and recall values improved across all datasets from these two platforms. In contrast, results from platform X showed a slight average decrease of 1.8 in F1 and 3.9 in recall. 
Additionally, slight decreases in average precision (0.6–3.8), recall (1.5–6.3), and F1 (1.8–1.9) were observed for some datasets with unchanged schemes. These decreases may be attributable to the introduction of new feedback category definitions, as the original definitions were dataset-specific.
On the Other hands, DS2, which underwent only minor changes in its classification scheme, and DS3, where the predefined categories remained unchanged, nevertheless showed improvements in F1 and recall. These improvements can be attributed to the use of more comprehensive definitions in the request prompts, which clarified category descriptions.

Among the datasets, DS1, DS4, and DS5, which contained a larger number of categories, showed the most substantial F1 improvements, with increases of 27.4, 9.2, and 7.8, respectively. This outcome was expected, as the new classification scheme introduced a higher level of abstraction. 
Our analysis of RQ1 revealed that the minimal distinctions between fine-grained categories often resulted in errors when identifying user feedback.
This finding indicates that LLMs face challenges in accurately predicting fine-grained categories.

The predictions from the studied LLMs also revealed reduced variability across models under the coarse-grained scheme. Specifically, the average F1 difference decreased from 14.5 in the original schemes to 6.0 in the coarse-grained scheme. A similar reduction was observed for recall (from 15.2 to 6.4) and precision (from 7.2 to 5.6). While the maximum values (often associated with GPT-4o) for unchanged-scheme datasets showed a slight decline, the minimum values (often associated with Flan-T5) improved across nearly all datasets, except DS8. This demonstrates that applying LLMs with a coarse-grained classification scheme can help ensure a minimum level of accuracy across different models.

\begin{figure*}[!t]
    \centering
    \captionsetup[subfloat]{labelfont=footnotesize ,textfont=footnotesize}
    \subfloat[F1 comparison]{\includegraphics[width=1.75in]{pics/average_F1_1.jpg}
    \label{fig:rq1_average_f1}}
    \hfil
    \subfloat[Precision comparison]{\includegraphics[width=1.75in]{pics/average_precision_1.jpg}
    \label{fig:rq1_average_precision}}
    \hfil
    \subfloat[Recall comparison]{\includegraphics[width=1.75in]{pics/average_recall_1.jpg}
    \label{fig:rq1_average_recall}}
    \caption{Comparison of LLM results between the original and coarse-grained classification schemes across precision, recall, and F1.}
\label{fig:rq1_shaded_figures}
\end{figure*}

\section{Proposed Two-tier LLM Routing Strategy}
\label{sec:proposed}
Our empirical study demonstrates that employing LLMs for feedback classification can substantially improve prediction accuracy. However, high-capacity LLMs demand considerable computational resources and incur significant costs of commercial models, making them impractical for large-scale deployment, particularly for popular applications that receive thousands of feedback entries daily.
In contrast, fine-tuned transformer-based models (like BERT~\cite{Bert_model}, DistilBERT~\cite{Distilbert_model}, and RoBERTa~\cite{Roberta_model}) have achieved state-of-the-art results in prior work on feedback classification~\cite{Hadi}. These models are lighter and more efficient for processing large volumes of user feedback compared to LLMs. Nonetheless, they still contain millions of parameters, and their fine-tuning requires substantial amounts of labeled data. When training data is limited, as is often the case with existing labeled feedback datasets, their accuracy can degrade considerably.

\begin{figure}
\centering
  \includegraphics[width=4in]{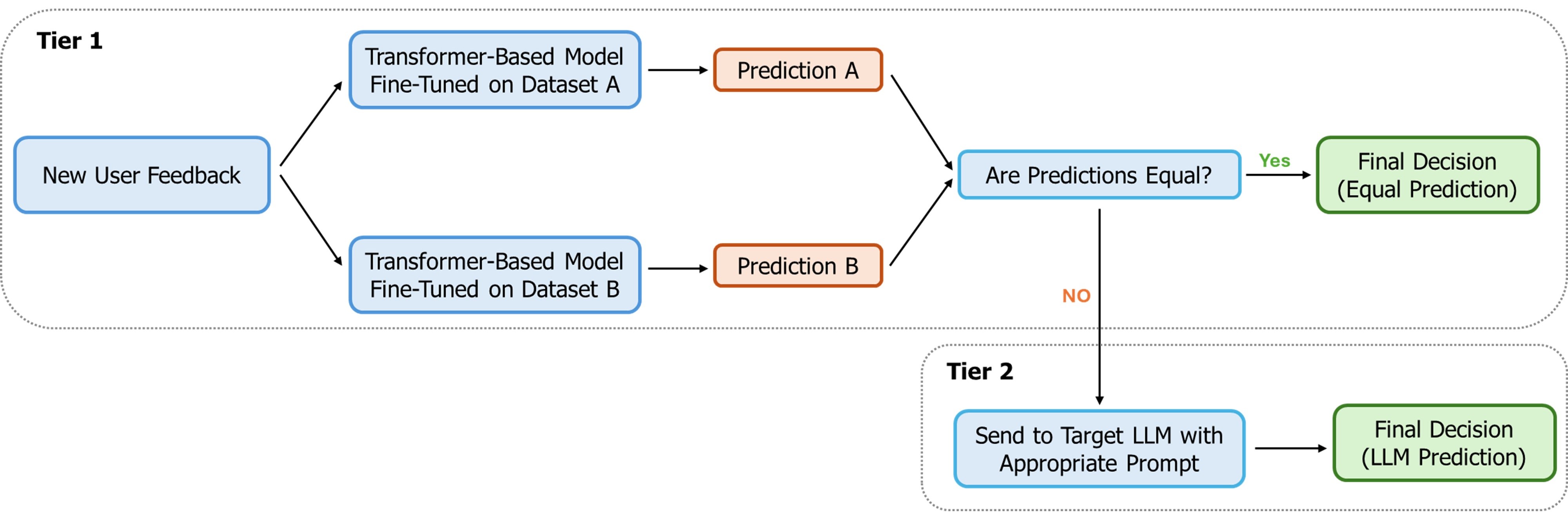}
\caption{Proposed two-tier LLM routing strategy for user feedback classification}

\label{fig:routing_strategy}  
\end{figure}

To address these challenges, this paper introduces a two-tier routing strategy designed to balance efficiency and accuracy (\figref{routing_strategy}). In the first tier, two fine-tuned models, trained on manually labeled datasets, are employed to classify new feedback instances. If their predictions converge, the consensus label is accepted as the final decision. If they disagree, the instance is escalated to the second tier. In the second tier, ambiguous feedback instances are routed to a high-capacity LLM, which is prompted with carefully designed instructions to conduct a more nuanced classification. The strategy allocates resources to complex cases, while simple instances are processed efficiently in the first tier.

\section{Experimental Evaluation of Routing Strategy(RQ2)}
\label{sec:rq2}
To address our second research question, we conducted an experiment to compare the accuracy and efficiency of feedback classification achieved by our proposed two-tier routing strategy against several state-of-the-art approaches. The objective is to demonstrate that leveraging LLMs selectively through the routing strategy can substantially improve efficiency while preserving accuracy. 

In recent years, several approaches have been proposed to improve the accuracy of user feedback classification, which serve as baselines for this experiment. Recent work~\cite{Abedini} has shown that augmenting training datasets, particularly with app-specific data, improves classifier accuracy. Similarly, Devin et al.~\cite{Devin} demonstrated that merging manually labeled datasets increases classification accuracy. Other studies~\cite{He,Moller_1,Ding} have employed LLMs as annotators to enrich existing manually labeled datasets, thereby addressing the bottleneck of scarce labeled data. 
Based on these insights, we considered the following baselines in this experiment:

\begin{itemize}
    \item \textbf{Fine-Tuned Model}: A feedback classification model that is fine-tuned on a manually labeled dataset.
    \item \textbf{Zero-Shot Model}: A target LLM are applied in a zero-shot setting for feedback classification task. 
    \item \textbf{Feedback-Augmented Model}: A feedback classification model that is fine-tuned on a manually labeled dataset augmented with another manually labeled dataset.
    \item \textbf{Random-Augmented Model}: A feedback classification model  that is fine-tuned on a manually labeled dataset augmented with a randomly selected LLM-labeled dataset.
    \item \textbf{App-Specific Augmented Model}: A feedback classification model  that is fine-tuned on a manually labeled dataset augmented with a randomly selected app-specific LLM-labeled dataset.
\end{itemize}

In this experiment, we focus on app store datasets that contain sufficient manually labeled samples, enabling us to apply various baseline approaches for a comprehensive analysis. Since DS1 explicitly identifies the app associated with each feedback, we use it as the truth dataset for evaluating classification approaches. DS1 includes feedback for seven applications; due to space and resource constraints, we selected three representative apps, Dropbox, WhatsApp, and Pinterest, as target applications for this experiment. For training the fine-tuned models in the first tier of our routing strategy, we use the DS3 and DS4 datasets because they provide relatively balanced category distributions suitable for training, and we employ GPT-4o in the second tier because it yielded the best results in RQ1. In addition, the DS2, DS3, and DS4 datasets contribute to the manually labeled portion of the training data used for the baseline models. To enable a comprehensive evaluation under various conditions, we define the following experimental settings:

\begin{itemize}
    \item \textbf{DS2–DS3 Setting}: DS2 and DS3 serve as the manually labeled dataset in the corresponding baselines, while DS4 is used as the augmentation dataset in the feedback-augmented model.
    \item \textbf{DS2–DS4 Setting}: DS2 and DS4 serve as the manually labeled dataset in the corresponding baselines, while DS3 is used as the augmentation dataset in the feedback-augmented model.
    \item \textbf{DS3–DS4 Setting}: DS3 and DS4 serve as the manually labeled dataset in the corresponding baselines. Because DS2 contains a limited number of samples, it cannot be used as an augmentation dataset. Therefore, the feedback-augmented model is not employed in this setting.
    \item \textbf{DS2–DS3–DS4 Setting}: DS2, DS3 and DS4 serve as the manually labeled dataset in the corresponding baselines. As no additional manually labeled dataset is available, the feedback-augmented model is not employed in this setting. 
\end{itemize}

Given the more effective utilization of LLM capabilities in the coarse-grained classification scheme (\secref{rq1_experiment2}), which includes three main categories, \emph{bug report}, \emph{feature request}, and \emph{other}, we conducted this experiment using this scheme. Based on the results, although applying LLMs with a coarse-grained scheme can enhance the accuracy of user feedback classification, they did not achieve sufficient precision to generate new labeled datasets for the random-augmented and app-specific augmented baselines. The precision metric, representing the proportion of correct predictions for user feedback categories, ranged between 52.2 and 88.9 across the datasets. Consequently, the samples generated by the LLMs do not provide adequate accuracy to create labeled datasets that could reliably support the training of a feedback classification model.

To address this issue, we propose using datasets labeled via the consensus of multiple LLM predictions to create high-accuracy LLM-labeled datasets. The~\tabref{RQ2_results_unified} presents results for samples that were assigned the same label across our different LLMs. We also provide results for the forum and X datasets for further comparison. The best results correspond to the app store datasets, except for DS2, with an average F1 of 91.8, precision of 93.8, and recall of 90.3. Generally, higher precision indicates more reliable predictions, while higher recall suggests better identification of diverse samples. Given the relatively high precision and recall in app store datasets, we conclude that LLM-generated unified labels can be valuable for fine-tuning lighter models. Since different LLMs were trained on varied datasets, their consensus in labeling samples can lead to reliable predictions.

\begin{table*}
    \caption{Results of LLMs consensus for feedback classification based on the coarse-grained scheme in a zero-shot setting}
    \centering
    \setlength{\tabcolsep}{6pt}
    \makebox[\textwidth][c]{
    \begin{tabular}{|c|ccccc|ccc|ccc|}
        \hline       
            \multirow{2}{*}{\textbf{Metric}}
            & 
            \multicolumn{5}{|c|}{\textbf{App Stores}}
            &  \multicolumn{3}{|c|}{\textbf{Forums}} 
            &  \multicolumn{3}{|c|}{\textbf{X}} 
            \\
            \cline{2-12}
            &
            DS1 & DS2 & DS3 & DS4 & Average &
            DS5 & DS6 & Average &
            DS7 & DS8 & Average\\
        \hline
        P
        &
        93.5 & 57.2 & 91.2 & 95.7 & 84.4 &
        61.2 & 83.1 & 72.2 &
        91.9 & 67.0 & 79.5
        \\
        \cline{1-12}
        R
        &
        91.7 & 62.2 & 88.0 & 93.1 & 83.8 &
        84.6 & 83.4 & 84.0 &
        92.2 & 58.8 & 75.5
    	\\
    	\cline{1-12}
        F1
        &
        92.0 & 55.6 & 89.4 & 94.3 & 82.8 &
        63.4 & 83.1 & 73.3 &
        92.0 & 59.0 & 75.5
	    \\
	   \hline
    \end{tabular}
    }
\label{table:RQ2_results_unified}
\end{table*}

Therefore, we apply this method to generate an LLM-labeled dataset of user feedback, which is subsequently used to augment the training dataset for the random-augmented and app-specific augmented baselines. The augmented datasets are created by integrating LLM-labeled and manually labeled datasets to fine-tune the feedback classification models. This augmentation enables us to examine the impact of increased training dataset volume on model accuracy. To ensure consistency in our result analysis, we maintain the same ratio of LLM-labeled dataset volume to manually labeled dataset volume across all categories. Given the limited number of samples in our target datasets, we follow prior work~\cite{Abedini} and set the augmentation ratio at 0.3.

To enable a more comprehensive analysis, we randomly selected user feedback to construct the LLM-labeled dataset from the general and app-specific datasets for the random-augmented and app-apecific augmented baselines, respectively. The general dataset consists of over 2.5M feedback from 1,132 apps across 29 non-gaming categories. The app-specific datasets include over 100K recent user feedback from selected target apps in Google Play that collect using the google-play-scraper library~\cite{google-play-scraper}. On the other hand, we also compare the cost of the proposed routing strategy with that of other baselines. In this experiment, cost is measured by the total number of requests sent to the LLM (\emph{request count}) and the total number of tokens processed (\emph{token count}). For simplicity, we exclude the computational and time costs associated with fine-tuning the classification models.

\subsection{Results}

As shown in \tabref{rq2_accuracy}, our routing strategy delivers competitive performance compared to the zero-shot setting, with only a negligible difference in average F1: 1.5 lower for bug reports and a 0.3 improvement for feature requests. This corresponds to retaining 98.4\% and 100.4\% of the zero-shot model’s accuracy, respectively.
In some cases, such as the Pinterest app for bug reports and the Dropbox and WhatsApp apps for feature requests, our approach even outperforms the zero-shot baseline. The improvements for Dropbox and WhatsApp are particularly notable, with gains of 4.9 and 8.3, respectively. These outcomes are reasonable, as in our proposed strategy, 95.1\% of bug report instances and 96.0\% of feature request instances resolved in the first tier produce the same classification outcome as if they had been sent directly to the LLM.

Additionally, within each setting, the fine-tuned models yield the lowest F1 scores, underscoring the limited generalizability of manually labeled datasets. Augmenting the training dataset with additional labeled data improves accuracy across all settings. In particular, incorporating LLM-labeled datasets consistently boosts accuracy compared to fine-tuned and feedback-augmented baselines. Using app-specific feedback for augmentation does not always yield better results than leveraging general feedback. A plausible explanation is that the general feedback dataset covers a wider variety of cases, whereas the app-specific dataset is more restricted and reflects only the most recent app-related feedback. Despite these improvements, baseline models still have a substantial accuracy gap compared to the zero-shot setting.

On the other hand, our approach incurs significantly lower costs compared to other LLM-based approaches. \tabref{rq2_accuracy} summarizes the \emph{request count} and \emph{token coout} for approaches that leverage LLMs to improve performance. As shown, approaches that use LLMs for dataset augmentation result in the highest costs, although these are one-time expenses limited to preparing the labeled dataset. In contrast, our proposed strategy achieves the lowest ongoing cost, with 375 requests and 10,833 tokens, corresponding to reductions of 67.8\% and 66.3\% in cost compared to the zero-shot setting, respectively. These findings highlight the potential of hybrid LLM routing to bridge the gap between high-performing but expensive zero-shot models and lower-cost, less generalizable fine-tuned models.


\begin{table*}
    \caption{F1 scores of baseline models and the proposed routing strategy across all settings. Abbreviations: FT = fine-tuned model, Feed-Aug = Feedback-Augmented, Ran-Aug = Random-Augmented, App-Aug = App-Specific Augmented, TTRS = Two-Tier Routing Strategy. The suffix DS\# indicates the experimental setting.}
    \centering
    \setlength{\tabcolsep}{1pt}
    \makebox[\textwidth][c]{
    \begin{tabular}{|l|c|c|c|c|c|c|c|c|}
        \hline       
            \multirow{2}{*}{\textbf{Approach}}
            &  \multicolumn{4}{|c|}{\textbf{Bug Report}}
            &  \multicolumn{4}{|c|}{\textbf{Feature Request}} 
            \\
            \cline{2-9}
            &
            Dropbox & Pinterest & WhatsApp & Average &
            Dropbox & Pinterest & WhatsApp & Average
            \\
        \hline
        FT-DS23
        &
        46.4 & 33.0 & 24.6 & 34.7 & 34.2 & 30.5 & 34.9 & 33.2\\
    	\hline
        Feed-Aug-DS23
        &
        74.2 & 69.4 & 64.2 & 69.3 & 61.9 & 62.5 & 64.2 & 62.9\\
    	\hline
        Ran-Aug-DS23
        &
        77.9 & 75.6 & 76.1 & 76.5 & 56.0 & 58.1 & 62.6 & 58.9\\
    	\hline
        App-Aug-DS23
        &
        81.0 & 77.9 & 79.8 & 79.6 & 56.0 & 46.4 & 60.5 & 54.3\\
    	\thickhline
        FT-DS24
        &
        76.2 & 71.9 & 71.6 & 73.2 & 52.6 & 51.9 & 54.1 & 52.9\\
    	\hline	
        Feed-Aug-DS24
        &
        84.1 & 84.5 & 82.1 & 83.6 & 48.3 & 39.6 & 39.8 & 42.6\\
    	\hline
        Ran-Aug-DS24
        &
        84.0 & 86.0 & 84.9 & 85.0 & 59.0 & 57.9 & 56.8 & 57.9\\
    	\thickhline
    	FT-DS34
        & 85.3 & 89.7 & 82.8 & 85.9 & 71.6 & 67.4 & 67.1 & 68.7\\
    	\hline	
        Ran-Aug-DS34
        &  86.3 & 91.3 & 88.0 & 88.5  & 71.2 & 65.4 & 67.8 & 68.2\\
    	\hline
        App-Aug-DS34 & 85.5 & 90.4 & 86.0 & 87.3 & 71.9 & 65.4 & 67.2 & 68.2 \\
    	\thickhline
        FT-DS234
        &
        84.3 & 85.8 & 74.0 & 81.4 & 54.1 & 41.3 & 43.1 & 46.2\\
    	\hline	
        Ran-Aug-DS234
        &
        82.2 & 90.8 & 87.4 & 87.4 & 64.0 & 58.8 & 54.5 & 59.1\\
    	\hline
        App-Aug-DS234
        &
        85.4 & 84.2 & 85.4 & 85.0 & 68.1 & 63.7 & 46.6 & 59.5\\
    	\thickhline
        \textbf{Zero-Shot}
        &
        \textbf{92.7} & \textbf{92.9} & \textbf{93.1} & \textbf{92.9} & \textbf{74.3} & \textbf{84.8} & \textbf{72.2} & \textbf{77.1}\\
    	\hline
    	\textbf{TTRS}
        &
        \textbf{88.5} & \textbf{95.4} & \textbf{90.4} & \textbf{91.4} & \textbf{79.2} & \textbf{72.4} & \textbf{80.5} & \textbf{77.4}\\
        \hline
    \end{tabular}
    }
\label{table:rq2_accuracy}
\end{table*}

\begin{table*}
    \caption{Request count and token count of baseline models and the proposed routing strategy across all settings. Abbreviations: Ran-Aug = Random-Augmented, App-Aug = App-Specific Augmented, TTRS = Two-Tier Routing Strategy. The suffix DS\# indicates the experimental setting.}
    \centering
    \setlength{\tabcolsep}{1pt}
    \begin{tabular}{|>{\columncolor{gray!30}}c|c|c|>{\columncolor{gray!30}}c|c|c|}
        \hline
        \textbf{Approach} & 
        \textbf{Request Count} & 
        \textbf{Token Count} &
        \textbf{Approach} & 
        \textbf{Request Count} & 
        \textbf{Token Count}
        \\
        \hline
        Ran-Aug-DS23 & 
        1,351 & 30,395 &
        App-Aug-DS23 & 6,740 & 127,978
        \\
        \hline
        Ran-Aug-DS24 & 1,307 & 26,522 & 
        App-Aug-DS24 & 6,215 & 110,618
        \\
        \hline
        Ran-Aug-DS34 & 1,895 & 43,369 & 
        App-Aug-DS34 & 13,967 & 173,808
        \\
        \hline
        Ran-Aug-DS234 & 18,839 & 201,807 &
        App-Aug-DS234 & 2,276 & 48,646
        \\
        \hline
        Zero-Shot & 1,165 & 32,182 & 
        \textbf{TTRS} & \textbf{375} & \textbf{10,833}
        \\
        \hline
    \end{tabular}
\label{table:rq2_cost}
\end{table*}

\section{Experimental Evaluation of First-Tier Configurations (RQ3)}
\label{sec:rq3}
In the first tier of our routing strategy, feedback samples are classified by multiple fine-tuned models, with their consensus determining the final decision. To assess the impact of different model configurations, we conducted an experimental comparison of alternative setups for this layer. As demonstrated in RQ2, augmenting the manually labeled dataset with additional LLM-labeled data improves classification accuracy. Building on this result, we evaluated fine-tuned models trained on either the original labeled dataset or the randomly augmented version as first-tier decision makers. For generating the augmented data, In this experiment, we employed GPT-4o, which outperformed the other target LLMs in RQ1. \tabref{rq3_comparative_study} summarizes the average F1-scores achieved for the bug report and feature request categories across the different configurations.

Using the augmented datasets does not improve accuracy and reduces the average F1 scores in all cases except for the DS2–DS3–DS4 configuration. This reduction is primarily driven by lower accuracy in the feature request category. Specifically, while augmented datasets improve classification accuracy for bug reports, they decrease it for feature requests. This decline may be due to errors in identifying new instances labeled via LLM consensus, as bug reports typically contain more general feedback, whereas feature requests often include app-specific details that make correct classification more challenging. The difference between DS3–DS4 and Aug-DS3–DS4 is relatively small, making the augmented variant more cost-effective considering the additional effort required to generate the labeled dataset. In contrast, the DS2-based model shows lower accuracy due to class imbalance, resulting in reduced final accuracy when fine-tuned models based on this dataset are used in the first tier. Furthermore, combining three datasets leads to a noticeable drop in the average F1 of feedback classification. One possible explanation is that achieving consensus is more difficult in this setting, and in our dataset, most consensus cases occur on incorrect predictions.

\begin{table*}
    \caption{F1 scores of first-tier configurations in the experimental comparison}
    \centering
    \setlength{\tabcolsep}{3pt}
    \begin{tabular}{|>{\columncolor{gray!30}}c|C{1.2cm}|C{1.2cm}|C{1.2cm}|>{\columncolor{gray!30}}c|C{1.2cm}|C{1.2cm}|C{1.2cm}|}
        \hline
        \textbf{Config.} & 
        \textbf{Bug} & 
        \textbf{Feature} &
        \textbf{Average} &
        \textbf{Config.} & 
        \textbf{Bug} & 
        \textbf{Feature} & 
        \textbf{Average}
        \\
        \hline
        DS2-DS3 & 89.3 & 72.4 & 80.9 &
        Aug-DS2-DS3 & 90.4 & 70.9 & 80.7
        \\
        \hline
        DS2-DS4 & 83.0 & 77.7 & 80.4 &
        Aug-DS2-DS4 & 88.1 & 63.1 & 75.6
        \\
        \hline
        \textbf{DS3-DS4} & \textbf{91.4} & \textbf{77.4} & \textbf{84.4} &
        Aug-DS3-DS4 & 92.0 & 76.0 & 84.0
        \\
        \hline
        DS2-DS3-DS4 & 78.7 & 39.1 & 58.9 &
        Aug-DS2-DS3-DS4 & 86.8 & 41.5 & 64.2 
        \\
        \hline
    \end{tabular}
\label{table:rq3_comparative_study}
\end{table*}

\section{Discussion}
\label{sec:discussion}
Our experimental results demonstrate that LLMs can significantly enhance the accuracy of feedback classification across various datasets. Compared to fine-tuned models trained on manually labeled datasets, LLMs generally produce more accurate predictions in the zero-shot setting. However, they do not guarantee better identification of all sample types within each category. As such, in scenarios where identifying all sample types is critical, such as detecting bug reports, fine-tuned models may be a more suitable option.
LLMs also face challenges in predicting fine-grained categories, performing better when the categories are abstract, distinct, and well-defined.
Despite these benefits, the practical use of LLMs is constrained by their high computational demands, slower response times, and significant operational costs.
Applying zero-shot classification directly to the vast volume of user feedback generated daily, particularly for popular apps, is therefore not feasible. To address these limitations, we proposed a hybrid routing strategy that balances accuracy with efficiency.
By routing only complex instances to high-capacity LLMs, the two-tier structure maintains competitive accuracy while substantially reducing cost.
Our approach achieves accuracy comparable to zero-shot results while reducing request costs by 67.8\% and token costs by 66.3\%. In addition, it retains 98.4\% to 100.4\% of the average zero-shot F1 score for bug reports and feature requests, respectively, and even surpasses it in certain cases. Notably, more than 95\% of the examples resolved in the first tier match the classification outcome of the LLM, reinforcing the cost-effectiveness of this strategy.

Beyond routing, leveraging LLMs as annotators to augment manually labeled datasets emerges as another viable operational pathway. Our experiments indicate that while the accuracy of individual LLMs may not always suffice for producing high-quality labeled data, combining their consensus yields more reliable annotations, particularly for app store feedback. Interestingly, app-specific augmentation did not outperform general-purpose datasets, suggesting that broader datasets with diverse samples provide greater utility. This finding highlights the feasibility of building a general-purpose augmentation dataset for fine-tuning lighter models. Such models can deliver relatively accurate results at lower computational cost and faster response times.Although they may not reach the accuracy achieved by zero-shot LLMs or by our approach, they consistently outperform fine-tuned models trained solely on small manually labeled datasets, making them a practical option for real-world deployment.

\section{Threats to Validity}
\label{sec:threat}
One potential threat to \emph{external validity} in our study is the limited generalizability of the results. To mitigate this concern, we targeted manually labeled datasets from three distinct platforms, each following diverse classification schemes. In addition, we selected several LLMs known for their strong performance across a wide range of tasks to help ensure competitive outcomes.
A potential threat to \emph{internal validity} lies in the risk of incorrectly defining the primary classification schemes. To address this, we relied on the definitions provided in the original papers associated with the target datasets when designing the primary classification prompts. For categories related to coarse-grained classification schemes, we employed a combination of primary prompts to improve accuracy.
We also aimed to base our comparisons on the latest studies in this field. Accordingly, we adopted the approach proposed by Devine et al.~\cite{Devin} as one of the baselines for RQ3. For training the classification models, we selected the pre-trained DistillBERT model, which is based on the transformer architecture.
A threat to the \emph{construct validity} of our study involves potential human errors in implementation and result analysis. To reduce this risk, we conducted multiple thorough reviews of the code and made various artifacts from this study publicly available~\cite{replicationpackage}, thereby also increasing the \emph{reliability} of our results. This availability enables future research, replication, and evaluation by other researchers.

\section{Conclusions and Future Work}
\label{sec:conclusion}
Extensive research has been conducted on feedback classification using NLP techniques and machine learning algorithms. 
However, the generalization capability of models trained on existing labeled datasets remains unsatisfactory. Recent studies indicate that LLMs demonstrate competitive performance in various downstream tasks compared to state-of-the-art approaches.
In this study, we conducted a comprehensive evaluation of eight datasets extracted from three platforms, each featuring diverse classification schemes with varying levels of abstraction, to assess the predictive capabilities of different LLMs in the zero-shot setting. We focused on four powerful LLMs: GPT-3.5, GPT-4o, Llama3, and Flan-T5. 
Through this empirical study, we highlighted how scheme design influence predictive performance. Building on these insights, we proposed a two-tier routing strategy in which simple feedback instances are handled by lightweight fine-tuned models, while complex or ambiguous cases are escalated to high-capacity LLMs.
Experimental results demonstrate that our approach retains 98.4\% to 100.4\% of zero-shot LLM accuracy while reducing request and token costs by 67.8\% and 66.3\%, respectively. This strategy provides a scalable and practical solution for real-world applications, making it feasible to process large volumes of feedback efficiently without substantial loss in accuracy. 

Future work could expand this study by augmenting manually labeled datasets with a larger volume of LLM-labeled datasets and examining the impact of dataset size on performance improvements. Additionally, dataset augmentation could be approached in different ways, such as using LLM-labeled datasets to balance existing human-labeled datasets.


\bibliographystyle{ACM-Reference-Format}
\bibliography{references}

\end{document}